# Comunicación interactiva en salud y construcción de identidad en redes sociales de personas con baja visión

Gustavo Caran[1], Ronaldo Araujo[2], Críspulo Travieso Rodríguez[3]

[1] 0000-0002-1199-5002 Instituto Brasileiro de Informação em C&T, Brasil. gmcaran@gmail.com
[2] 0000-0003-0778-9561 Universidade Federal de Alagoas, Brasil. ronaldo.araujo@ichca.ufal.br
[3] 0000-0002-0774-0728 Universidad de Salamanca, España. ctravieso@usal.es

**Resumen**:
En el uso y apropiación del ámbito digital por parte del movimiento Bengala Verde en Brasil, el carácter informativo en salud es un recurso discursivo fundamental, y se desarrolla infundido por (auto)concepciones sobre los modos característicos de pensar, sentir y reaccionar de la persona con baja visión. El objetivo de este estudio es explorar el discurso de estas personas respecto a su propia identidad social en medios y redes sociales. Basado en la Comunicación Interactiva en Salud, se analizaron vídeos de YouTube y publicaciones en Facebook e Instagram del Grupo Virtual Stargardt, una comunidad virtual compuesta por personas de baja visión y que padecen la enfermedad de Stargardt. A partir de la selección de fragmentos de esos discursos y de su codificación, se identificaron 42 rasgos identificadores, agrupados en un modelo sintético de 8 categorías: (T1) tenemos baja visión y usamos un bastón verde; (T2) tenemos sensibilidad a la luz y usamos gafas de sol; (T3) tenemos dificultades en el día a día, pero hacemos uso de estrategias y tecnologías asistenciales; (T4) somos personas con discapacidad y la ley garantiza nuestros derechos; (T5) queremos saber sobre nuestra enfermedad y somos portavoces de la misma; (T6) tenemos una discapacidad visual pero somos más que eso; (T7) convivimos con ciertos problemas y contradicciones por la forma en que vemos las cosas, y; (T8) nos reconocemos como comunidad. Estos rasgos se construyen de forma interactiva, por medio de una elaboración pautada en momentos de acuerdo, discordancia y sentido del humor. Los discursos contribuyen a la consolidación de una identidad propia, distinta a la de la persona invidente pero similar en cuanto al marco legal de la persona con discapacidad visual.

**Palabras clave:** Persona com discapacidad; Baja visión; Identidad; Comunicación interactiva em salud; Redes sociales.

**Resumo**
No uso e apropriação dos ambientes digitais pelo movimento Bengala Verde no Brasil, o caráter informativo em saúde é um recurso discursivo central, e ocorre imbuído de (auto) concepções a respeito de modos característicos de pensar, sentir e agir da pessoa com baixa visão. O objetivo deste trabalho foi explorar o discurso da pessoa com baixa visão nas mídias sociais sobre sua própria identidade social. Apoiada na Comunicação Interativa em Saúde, foram investigados vídeos do Youtube e postagens do Facebook e Instagram do Grupo Virtual Stargardt, uma comunidade virtual composta por pessoas com baixa visão e acometidas pela Doença de Stargardt. A partir da seleção de trechos e da sua codificação, foram identificados 42 traços identitários, agrupados em um modelo sintético com 08 categorias: (T1) somos baixa visão e o usamos a bengala verde; (T2) temos sensibilidade à luz e usamos óculos de sol; (T3) temos dificuldades no dia-a-dia, mas utilizamos estratégias e tecnologias assistivas; (T4) somos uma PcD e temos direitos garantidos por lei; (T5) buscamos entender sobre nossa doença e somos o porta-voz dela; (T6) temos uma deficiência visual, mas não somos apenas isso; (T7) convivemos com conflitos e contradições pela maneira como enxergamos as coisas, e; (T8) nós reconhecemos como uma comunidade. Tais traços são construídos interativamente, por meio de uma construção pautada em momentos de concordância, discordância e bom-humor. Os discursos se orientam na consolidação de uma identidade própria, distinta da pessoa com cegueira, mas semelhante no enquadramento legal da pessoa com deficiência visual.

**Palavras-chave:** Pessoa com Deficiência; Baixa Visão; Identidade; Comunicação Interativa em Saúde; Redes Sociais na Internet.

## Introducción

En los últimos 20 años Internet se ha convertido en un espacio de movilización social en diversos ámbitos. Conceptos como colaboración, participación, compromiso o cooperación se han incorporado a la web, modificando su configuración inicial, unidireccional, asíncrona y estática, hacia una comunicación cada vez más multidireccional, síncrona y dinámica (Recuero, 2009). La red se ha vuelto así un espacio de diálogo amplio y diverso, en el que las conversaciones se conforman y distribuyen mediante múltiples herramientas, como son Facebook, Twitter, Instagram, YouTube etc. (Recuero, 2012).

Por un lado, esta comunicación virtual es capaz de promover un entorno más democrático, en que se comparten y debaten perspectivas ideológicas y sociales distintas. Los grupos sociales minoritarios tienen ahora un espacio en el que pronunciarse, de defensa de su identidad y de sus posiciones políticas e ideológicas. Y por otro lado, estos espacios virtuales no están libres de la influencia del capital, de la globalización y de los vaivenes propios de una sociedad fluída, dinámica y condicionada de forma constante por impulsos de tipo emocional (Bauman & Mauro, 2016). Al estar supeditada a influencias masivas y de carácter heterogéneo, la web ha devenido en un espacio determinado por los distintos movimientos sociales, que articulan todo un entramado de esperanzas y sentimientos de indignación (Castells, 2017).

En este contexto, los grupos sociales minoritarios se movilizan para dejar su impronta en esa comunicación en red. Como una forma de lucha y de poder para su inclusión social, dichos grupos buscan un "lugar en la mesa de debate", manifestando no solo aquello que defienden en cuanto a posición político-ideológica, sino también construyendo, defendiendo y dando a conocer su identidad – además de armarse de las prácticas informacionales que contribuyen a este proceso -.

Dentro de los estudios de mediación de la información en el ámbito de la salud, la Comunicación Interactiva en Salud (CIS) se define a partir del uso que las personas hacen de los dispositivos informáticos para buscar, obtener y compartir información sobre salud (Araújo, Silva & Mota, 2015).

Para Eng et al., (1999), se considera CIS cualquier interacción de un individuo – consumidor, paciente, cuidador o profesional – mediante un dispositivo electrónico o tecnología de comunicación para acceder o transmitir información sanitaria o para recibir orientación o ayuda en un asunto relacionado con la salud. Las investigaciones previas sobre CIS se han ocupado más de la educación y la promoción de hábitos saludables (Eng & Gustafson, 1999; Murray et al, 2005; Rada, 2005) y solo algunas se han centrado en el análisis de la subjetividad y los rasgos identificadores de los sujetos durante el proceso de mediación e intercambio de informaciones suministradas por los medios digitales.

En cuanto a las intervenciones en redes sociales para promover la igualdad en salud, Welch et al. (2016) defienden que los distintos medios sociales están determinados en base a que concentran bloques de construcción funcional a partir del grado de interacción y comunicación entre los usuarios y los autores; uno de esos bloques es precisamente el de la identidad, definida en los términos en que los usuarios y los grupos se revelan y relacionan. Redundando en la idea planteada, ello cobra aún mayor relevancia en determinados colectivos; en palabras de Vaca Vaca et al (2012), "esta relación dialéctica entre interpretaciones y actos, entre individuos y contextos sociales, hace que el estudio de las representaciones sociales sea de utilidad en la comprensión de problemáticas comunitarias, como lo es la de la discapacidad, pues permite comprender cogniciones y sistemas explicativos que median las interacciones en determinados grupos".

Entre los numerosos grupos sociales minoritarios con presencia en internet, este trabajo se centra en las Personas con Discapacidad Visual (PDV), un grupo heterogéneo en cuanto a sus capacidades visuales y a sus rasgos socio-identificadores. Igual que los demás subgrupos de Personas con Discapacidad (PcD), las PDV buscan, también en la web, acceso a productos, servicios, informaciones y derechos. Términos como accesibilidad, inclusión social y digital, integración social y diseño universal representan ese esfuerzo promovido por las propias personas con discapacidad visual, por la sociedad y por los organismos públicos en construir una relación bidireccional entre este grupo social y la sociedad en que se inserta (García, 2102). Bidireccional en el sentido en que permite el acceso y la inclusión, facilitando que el individuo asuma plenas capacidades a la hora de interaccionar activa y pasivamente con la sociedad, logre una comprensión del mundo y de su posición social y constituya su identidad por medio de elementos materiales y simbólicos.

Tanto dentro como fuera del ámbito de la Ciencia de la Información (CI), varios trabajos han explorado el acceso y la accesibilidad en el contexto de las personas con discapacidad visual, discutiendo los factores actitudinales, técnico/tecnológicos, políticos y socioculturales que marcan su inclusión en el ámbito digital. Sin embargo, especialmente en CI, se ha abordado de forma escasa esa expresión identitaria de las PDV en internet, a través de los contenidos multimedia compartidos por ellas mismas en entornos en línea. ¿Cómo construye una persona con discapacidad visual su presencia digital? ¿Qué discurso defiende? ¿Qué rasgos identificadores manifiesta? ¿Y ante qué otros aspectos se posiciona de forma contraria? Estas preguntas motivaron que se llevara a cabo la siguiente investigación, cuyo objetivo es analizar de manera cualitativa, exploratoria y descriptiva la presencia digital y el discurso de las personas con discapacidad visual en la construcción de su identidad.

**2 Metodología**

Según las directrices de la Organización Mundial de la Salud (2003) los tipos de discapacidad visual pueden clasificarse en dos grupos: ceguera; y baja visión o visión insuficiente. Una persona invidente puede o no tener resto visual, pero en todo caso no lo usa de manera significativa para localizar objetos u orientarse. Frente a ello, la persona con baja visión usa ese remanente visual que posee en sus actividades cotidianas (Caran, 2013). Los parámetros clínicos de medición de la capacidad visual se han materializado en la creación de una clasificación normativa. Sin embargo, la experiencia de no ver, parcial o completamente, depende del contexto práctico, social y cultural. Según establecía en 2015 el nuevo *Estatuto da Pessoa com Deficiência*: la discapacidad se expresa en base a la desigualdad de oportunidades experimentadas (Dhanda, 200'8; Lago Júnior & Barbosa, 2016).

La distinción entre esos dos grupos se ha vuelto actualmente más evidente en Brasil, con la aparición online de movimientos sociales como Bengala Verde[1]. Sustentados en el lema *queremos que você nos veja*, el objetivo de este movimiento es difundir socialmente la identidad de la persona con baja visión, utilizando también el bastón para indicar que se es una PDV, pero diferenciándose de la persona invidente por el color verde. Esta iniciativa se inspiró en el proyecto social del mismo nombre, puesto en marcha en Argentina en 2012 por la profesora Perla Mayo. En Brasil, el movimiento *Bengala Verde* emplea internet para promover espacios de interacción entre personas con baja visión y para concienciar socialmente sobre su distinción con respecto a las personas que padecen ceguera.

---
[1] http://www.bengalaverde.com.br/ (Bastón verde)

En el contexto descrito, este trabajo analiza la presencia digital y los discursos de las personas con baja visión. Tras localizar los registros micro-documentales de los medios y redes sociales (Jeanneret, 2015), se comenzó un tratamiento metodológico de exploración inspirado en la teoría Actor-Red de Bruno Latour (2012). Las huellas digitales de los autores sobre los discursos de las personas con baja visión fueron monitorizados, empezando por la página web del movimiento *Bengala Verde* y ampliando la investigación a partir de la navegación por los hiperenlaces y referencias publicados desde allí.

En este examen se ha contemplado la identificación de comunidades online relevantes impulsadas por los propios usuarios (*User-Driven Social Support*) – esto es, orientadas a personas con baja visión y gestionadas por ellas mismas (Ahmed *et al.*, 2017) -. Tras esa fase inicial, se seleccionó el Grupo Virtual Stargardt (GVS)[2], creado en 2012 por los afectados por la enfermedad de Stargardt (ES), una patología que provoca la degeneración de los fotorreceptores de la zona central de la retina (mácula) y que lleva a la condición de baja visión. El GVS tiene presencia en Facebook, YouTube, Instagram y Whatsapp y se trata de una comunidad cuyo objetivo es compartir informaciones y apoyo social a los pacientes de esta dolencia.

Aunque la enfermedad de Stargardt está catalogada como enfermedad rara, es una de las patologías degenerativas de la retina más frecuentes en la población mundial, afectando a una de cada 10.000 personas. La perdida de capacidad visual (agudeza visual) se puede manifestar en la infancia o ya en la edad adulta. La facultad de ver detalles en la visión central se va reduciendo gradualmente, pero no suele derivar en ceguera. Por eso, la persona que padece esta enfermedad suele vivir como una persona con visión normal, si bien esta va perdiendo agudeza, hasta el momento en que se reconoce a sí misma y es legalmente reconocida como persona con baja visión (Retina Brasil, 2019).

De los canales del GVS se seleccionaron solo los de acceso público, concretamente su página en Facebook, su perfil en Instagram y su canal en YouTube. Se consideraron productos de comunicación interactiva en salud, y por tanto se analizaron como tales los contenidos de las entradas y comentarios. Se recogieron todos los vídeos de YouTube y publicaciones de Instagram, con sus correspondientes descripciones y reacciones. En el caso de Facebook, debido al gran volumen de publicaciones existentes, se optó por recopilar las publicaciones de 2018 y sus comentarios. Se usó la aplicación Netvizz[3] para la recogida automática de los datos de Facebook, mientras que en el resto de medios se procedió de manera manual. Este proceso se llevó a cabo el 5 de enero de 2019 y se recogió un total de 42 vídeos de YouTube, 18 publicaciones de Instagram y 70 *posts* de Facebook. Para el tratamiento de los datos se utilizó *Excel 365* y el análisis de los mismos se realizó con el programa ATLAS.ti. La figura 1 ilustra los pasos seguidos en la metodología de este trabajo.

Figura 1: Fases de la metodología empleada.

---

[2] http://www.stargardt.com.br/
[3] https://apps.facebook.com/107036545989762/?ref=br_rs

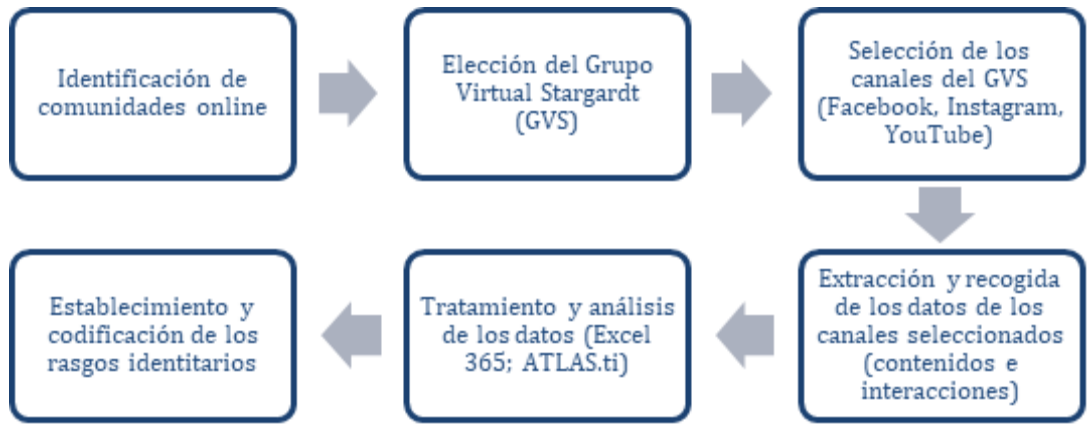

En los estudios sobre medios sociales, incluso en los orientados hacia el marketing digital, cada vez es más habitual el estudio de reglas de conducta aplicadas al análisis del comportamiento de los usuarios. Una de estas normas, esgrimidas por Scott (2013), destaca que, además del hecho de que los actores en redes sociales estén dispuestos e interesados en la interacción con otros y en la participación, las publicaciones dicen mucho sobre quienes las comparten. Dicho autor llega a afirmar que "usted es lo que publica". En esa línea, para esta investigación se partió de la premisa de que los contenidos de las publicaciones y de sus reacciones eran significativas sobre el grupo analizado, y los rasgos identificadores percibidos en las publicaciones fueron agrupados y descritos en función de sus características y similitudes.

Como se ha comentado, el análisis de los datos se realizó en ATLAS.ti, siguiendo las siguientes etapas: 1) importación de los vídeos de YouTube (42 documentos) y de las publicaciones y comentarios de Facebook (70 documentos) en formato *.pdf*; 2) importación de páginas web y vídeos compartidos en publicaciones de Facebook (26 documentos); 3) detección y citación de fragmentos de documentos que incluían percepciones o creencias sobre la persona con enfermedad de Stargardt, en expresiones como "yo soy…", "nosotros somos…", "la persona con ES es…", etc.; 4) agrupación y clasificación de la citaciones obtenidas en expresiones (códigos) que representaran los rasgos identificadores presentes en los contenidos compartidos, y; 5) análisis de los códigos y citas, ajustados y sistematizados según un modelo de síntesis.

**3 Resultados y discusión**

Los contenidos procedentes de vídeos de YouTube y las publicaciones de Facebook e Instagram permitieron la identificación de 42 códigos, que fueron agrupados en 8 rasgos identificadores de la persona afectada por la enfermedad de Stargardt. Estos rasgos están relacionados con la dimensión de los sentidos en la mediación de información (Marteleto, 2010), y su objeto de representación en construcción intersubjetiva es la persona con ES en su vida diaria respecto a su comunicación interactiva en salud. Estos rasgos representan tanto aspectos subjetivos, referidos a sus modos de pensar y sentir sobre sí mismo, como aspectos objetivos, sus formas de expresión y reacción en el uso de objetos cotidianos.

A la hora de llevar a cabo el análisis, este trabajo concebía que la identidad de la persona con ES no está separada del resto de su existencia. Lo que "ellos son" no esta separado del resto de "lo que viven" o de "lo que hacen" o "lo que usan" en el día a día. Por tanto, los rasgos identificadores son capas representativas híbridas sobre actores, acciones y artefactos de la vivencia de la persona con esta enfermedad, sin separar los sujetos y los objetos de su contexto social (Latour, 2012).

En la figura 2 se muestran los rasgos identificadores, indicando los distintos aspectos representados para cada uno de ellos. En concreto, dichos rasgos son los siguientes: (T1) tenemos baja visión y usamos un bastón verde; (T2) tenemos sensibilidad a la luz y usamos gafas de sol; (T3) tenemos dificultades en el día a día, pero hacemos uso de estrategias y tecnologías asistenciales; (T4) somos personas con discapacidad y la ley garantiza nuestros derechos; (T5) queremos saber sobre nuestra enfermedad y somos portavoces de la misma; (T6) tenemos una discapacidad visual pero somos más que eso; (T7) convivimos con ciertos problemas y contradicciones por la forma en que vemos las cosas, y; (T8) nos reconocemos como comunidad.

Figura 2: Rasgos identificadores de la persona con enfermedad de Stargardt

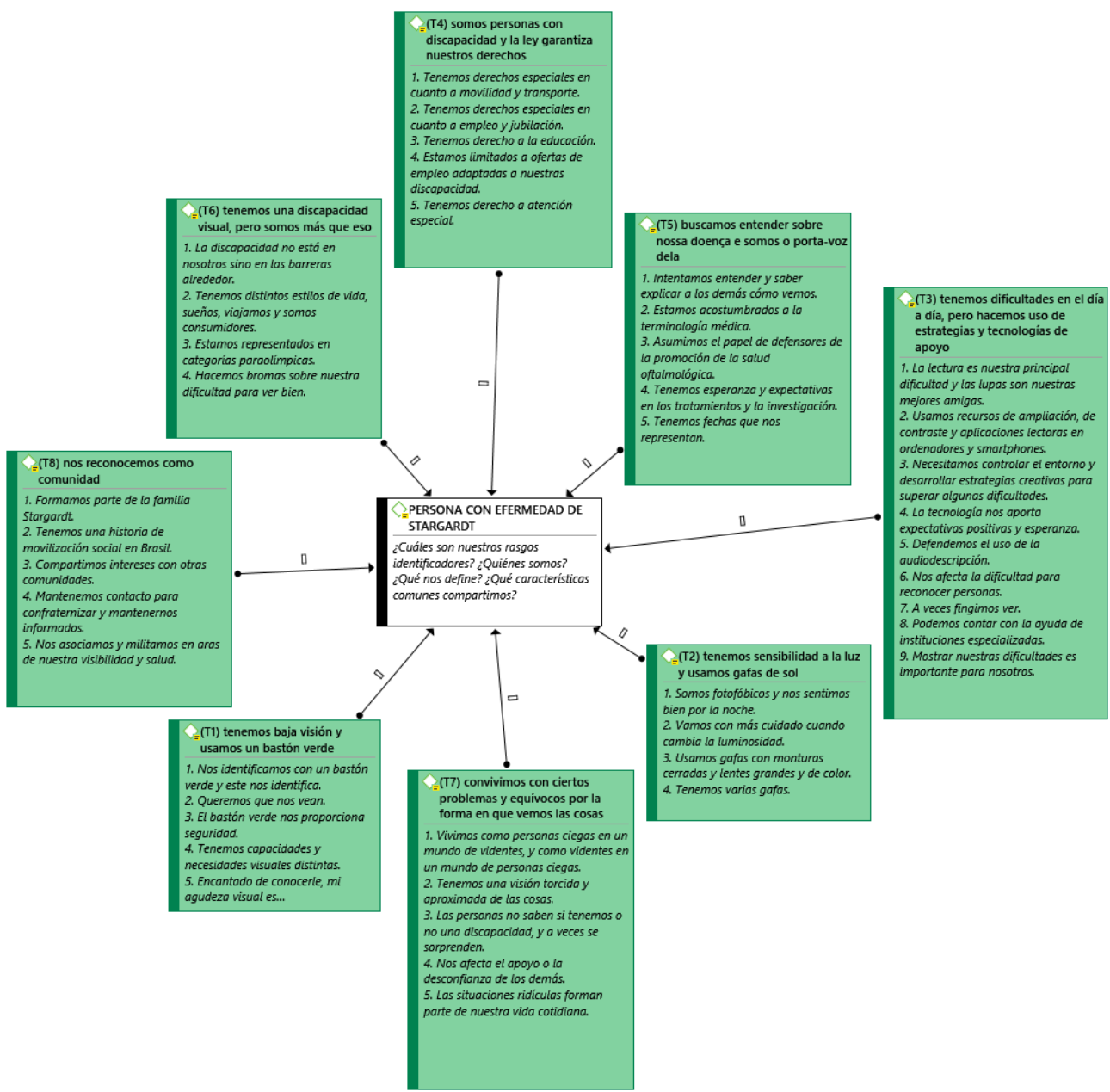

Los tres primeros rasgos asocian la identidad de la persona con ES a objetos que han sido incorporados a su rutina: el bastón, las gafas de sol y las tecnologías de apoyo. Los rasgos T4 y T5 representan el aspecto social recogido en la legislación y el papel de portavoces en la construcción social de la identidad de la persona con ES. Los rasgos T6 y T7 constituyen aspectos de la vida diaria que los definen más allá de la enfermedad (como estilo de vida), pero vinculados a los problemas y equívocos inherentes a la baja visión. Por último, el rasgo T8 se refiere a la relación entre las personas con ES, mediante su vinculación comunitaria e institucional.

**(T1) Tenemos baja visión y usamos un bastón verde**

El discurso más recurrente en el GVS demuestra una vinculación de la persona con enfermedad de Stargardt con la condición de discapacidad visual establecida por la Organización Mundial de la Salud (OMS), usando expresiones como "soy de baja visión", "tengo baja visión" y "tengo visión por debajo de lo normal". En una de las intervenciones se dice "[…] tengo enfermedad de Stargardt, así que soy subnormal" (Vídeo de YouTube de 9 de julio de 2016), apropiándose de esa definición normativa como aspecto inherente a esta enfermedad. En términos clínicos, no todas las personas con ES se consideran de baja visión (OMS, 2013). Sin embargo, con la evolución natural del proceso degenerativo de la visión (aún sin tratamiento para interrumpir o revertir este proceso), la persona con ES tiende a perder agudeza visual, y a encasillarse como de baja visión en algún momento de su vida.

Otro aspecto referido a la forma en que la persona con esta enfermedad se presenta a los demás es informar de su agudeza visual. Así se comprobó en el discurso de cinco individuos, según el siguiente fragmento: "me llamo Luciano Marques […], padezco la enfermedad de Stargardt […] y mi grado de visión es 20/100" (Vídeo de YouTube de 21 de enero de 2017). Esa asimilación de los patrones normativos de agudeza visual reconocidos por la OMS es una fracción cuyo numerador representa la distancia en que esa persona es capaz de ver un determinado objeto, siendo el denominador la distancia a la que lo ve una persona con visión normal. En otras palabras, tener una agudeza de 20/100 significa que esa persona tendrá que estar cinco veces más cerca para que sea capaz de ver el mismo objeto.

También estuvo presente en las declaraciones el incentivo del uso del bastón verde en la rutina de la persona con ES, relacionando este objeto con su identidad como persona con baja visión. El bastón blanco, como elemento simbólico reconocido socialmente como propio de las personas invidentes, se redimensiona aquí por la persona con baja visión por el color verde. Ello implica el mensaje de que "también somos personas con discapacidad, vemos con dificultad y tenemos necesidades distintas a las de los invidentes". Además de ser un instrumento que proporciona seguridad y mejor movilidad a estos individuos, se concibe como un objeto de señalización, para que los demás sepan que se trata de una persona con baja visión, tal como se muestra en la siguiente intervención:

> [Marina:] Lo importante del bastón verde es […] que las demás personas nos identifican. Porque puede que estés en un lugar en que vas a ver a alguien, pero puede que no lo veas. Esa persona te va a identificar como persona con discapacidad visual […] es importante, es importante porque te va a esquivar, no se va a tropezar contigo, porque estás usando el bastón. O si estás en algún lado y vas a pedir que alguien te ayude, en un banco o en una estación. Con el bastón, te identifican inmediatamente como discapacitado y esa persona te va a prestar auxilio, te va a ayudar. Y así no tienes que andar buscando a alguien que te eche una mano y pasar por más aprietos. (Vídeo de YouTube de 16 de febrero de 2018)

Este aspecto identificado en el presente estudio coincide plenamente con una investigación de enfoque cualitativo anterior, también centrada en personas con baja visión. En ella, los autores, Morel y Villalobos (2011) mantienen que lejos de negar la discapacidad, se empeñan en mostrarla y demostrarla, y recogen la afirmación de uno de los participantes que declara no temer "ponerse el cartel y usar un bastón sin importar si es blanco o verde, lo que importa es que lo necesitamos y con él nos sentimos seguros".

El lema "queremos que nos ayudéis" encarna la intención de persona con ES de que el color verde del bastón sea capaz de distinguirla en la sociedad. Esto es, promueve una construcción social de conocimiento y reconocimiento de sus características y necesidades propias. En los argumentos de Sony, una persona con baja visión que lleva dos años usando el bastón verde, se ilustra la importancia del color verde del bastón, al decir que "creo que […] desde el momento en que puedo identificarme con lo que me pasa, me siento más cómodo. Y las personas que me miran necesitan también saber lo que me pasa" (Post n.º 13 de Facebook).

**(T2) tenemos sensibilidad a la luz y usamos gafas de sol**

El uso de gafas de sol es también un elemento identificador de la persona con enfermedad de Stargardt, y es, ante todo, una consecuencia fisiológica de la propia dolencia: la elevada sensibilidad a la luz, denominada fotofobia. Según afirma Marina, "incluso en situaciones en que la claridad aparentemente no molesta, para nosotros puede ser excesiva […]" (Vídeo de YouTube de 12 de octubre de 2016), así que el uso de gafas oscuras no es solo una cuestión de cuidado de la salud de los ojos, sino un elemento imprescindible para el bienestar diario – incluso en entornos cerrados muy iluminados -.

Al igual que el bastón, las gafas de sol con monturas grandes que cubren todo el campo visual son elementos característicos de la persona con ES. Muchos poseen varias, con tipos de monturas y lentes adaptadas a distintos ambientes. Las molestias propias de la luminosidad diurna hacen que las noches sean el periodo más agradable y cómodo, según cuenta Luciano, un deportista habitual: "[…] mi visión es de 20/100. No es mucha, pero me es bastante para andar por la noche sin ningún problema". Y culmina afirmando que "[…] acabamos yendo en bici más por la noche. Me encanta montar en bici por la noche, es cuando más disfruto" (Vídeo de YouTube de 21 de enero de 2017).

**(T3) tenemos dificultades en el día a día, pero hacemos uso de estrategias y tecnologías de apoyo**

Uno de los principales problemas manifestados por las personas con esta enfermedad tiene que ver con la lectura, especialmente para aquellos que están en periodo de formación escolar o académica. La lectura de libros, pizarras, etiquetas de productos e incluso documentos escritos a mano se vuelven tareas difíciles a simple vista. Por ello, las lupas son una herramienta esencial para el día a día, ya sean ópticas o electrónicas. Esta relación con las lupas es tan estrecha que Marina (Vídeo de YouTube de 20 de agosto de 2016) afirma que son sus mejores amigas, que las usa a diario y que le son tan indispensables como las gafas de sol.

Los ordenadores y smartphones también son elementos cotidianos de la persona con ES, para cuyo uso han de emplear constantemente tecnologías de apoyo, como puedan ser las técnicas de ampliación de pantalla, el aumento del contraste y los lectores (conversores de texto a voz). Este tipo de aplicaciones se utilizan tanto al usar propiamente los ordenadores y teléfonos móviles, como al tomarlos como tecnologías de apoyo para la visualización de imágenes pequeñas o textos cortos. Por ejemplo, la cámara de un teléfono pude usarse para ampliar los textos impresos en un folleto o para ampliar un menú en una pantalla de un dispositivo electrónico que no esté adaptado a una persona con baja visión.

Aunque las diversas tecnologías de apoyo ya posibilitan ayudar en el día a día a estos individuos, aún no permiten la realización de ciertas prácticas de forma autónoma. Una de esas lagunas que todavía no se ha cubierto satisfactoriamente es la identificación de los rostros de personas – cuestión apuntada como fundamental por la persona con enfermedad de Stargardt -. Debido a la significativa reducción de la capacidad de ver los detalles, la persona con ES no percibe los rasgos faciales de los que le rodean. Esta característica innata de los seres humanos se considera primordial en el proceso de socialización, por lo que la pérdida de esta capacidad puede traer consecuencias prácticas y emocionales negativas en la relación interpersonal y en la autoestima. En palabras de Gabriela y Milka:

> [Gabriela:] Hoy día tenemos muchas aplicaciones que nos facilitan mucho las cosas, pero una de las que sigue siendo difícil para nosotros que tenemos una discapacidad visual es que no conseguimos disitiguir los rasgos faciales de la gente. Esto termina produciendo sufrimiento, muchas dificultades. (Vídeo de YouTube de 3 de marzo de 2016)
>
> [Milka:] *Ya no lograba reconocer a la gente, incluso aunque estuviesen bastante cerca de mí – a metro o metro y medio. Y esto fue mermando mucho mi autoestima y mi productividad. Llegué a cuestionarme mi capacidad para trabajar. (Vídeo de YouTube de 21 de junio de 2017)*

**(T4) somos personas con discapacidad y la ley garantiza nuestros derechos**

Las personas con esta enfermedad consideran que tienen una discapacidad, y por tanto sufren de forma colectiva las barreras a la accesibilidad y la desigualdad de oportunidades. Los acuerdos internacionales y la leyes nacionales, creadas y potenciadas especialmente en los últimos treinta años, atribuyen derechos especiales a las personas con discapacidad. Estas garantías legales cubren diversos aspectos de la vida de estas personas, como acceso a la educación, reservas en el mercado de trabajo, jubilaciones por invalidez, gratuidad y beneficios en el trasporte público, ventajas en la compra de vehículos, atención preferente y descuentos en diversos eventos. Todos estos aspectos se abordan con detalle en el GVS, lo que supone convertirse en un canal informativo y de incentivación al uso de estas prerrogativas legales.

Al recopilar toda la legislación que rige estos derechos especiales, el GVS establece el alcance y la definición legal de "qué es ser una persona con discapacidad y dónde nos encuadramos legalmente", "cuáles son las leyes que tratan de reducir nuestra desigualdad de oporunidades" y "qué derechos se aplican a nuestra discapacidad". Se conforma así una percepción de la persona con ES y sus derechos desde el punto de vista legal, señalando su importancia en la vida diaria, para facilitar su movilidad y desplazamientos, mantenerse económicamente o estudiar. Los carnets especiales para el transporte público y las plazas de aparcamiento reservadas son documentos que los legitiman como personas con algún tipo de discapacidad.

A pesar de que estas garantías legales persiguen la igualdad de oportunidades, la persona con enfermedad de Stargardt se encuentra muy limitada en cuanto a la reserva de plazas de empleo público. Según atestigua Milka (Vídeo de YouTube de 20 de mayo de 2017) la persona que padece

ES cuenta con total amparo legal para realizar los exámenes en condiciones especiales, pero siempre que pueda asumir cargos que sean compatibles con su discapacidad. Así, aún concursando a plazas por turno libre (no reservadas a personas con discapacidad), el candidato está circunscrito a aquellos puestos cuyas funciones se le consideren factibles.

**(T5) queremos saber sobre nuestra enfermedad y somos portavoces de la misma**

Igual que otras minorías, la persona que padece esta enfermedad actúa como portavoz de sus patrones de comportamiento, difundiendo información y defendiendo sus intereses (García, 2012). Este papel queda claramente patente en las intervenciones del GVS, en las que el carácter informativo sobre "cómo ve la persona con ES" (Vídeo de YouTube de 15 de noviembre de 2016), "qué es la enfermedad de Stargardt y cómo se manifiesta" (Vídeo de YouTube de 1 de abril de 2016), entre otros contenidos que explican a las personas que no tienen esta enfermedad qué significa padecerla. En estos discursos, aparecen explicaciones de conceptos técnicos de oftalmología, demostrando en la práctica una mayor familiaridad con la jerga médica.

A la vez que se subrayan las características biológicas y experimentales que contribuyen a la definición de la persona con ES, sus contenidos cumplen una función de promoción de la salud de los ojos y de divulgación científica, explicando aspectos fisiológicos de estos órganos y la importancia de determinados cuidados, como tomar el sol con filtros de protección UV-A y UV-B y la atención que ha de prestarse a ciertos síntomas, como el ardor o los ojos enrojecidos. El apoyo a la campaña *Abril Marrom* es un ejemplo de ello, donde el cuidado de la salud ocular asume una función simbólica vinculada a la deficiencia visual (Post n.º 15 de Facebook). Otras fechas destacadas, como el Día Internacional de las Personas con Discapacidad (3 de diciembre) o el Día Nacional de la Persona Ciega (13 de diciembre) y el Día Mundial de la Retina (25 de septiembre) también se difunden desde el GVS y son señaladas como especiales para la persona con ES, acercándolas conceptualmente a la noción de persona con discapacidad y a persona invidente, así como las vincula con la retina como parte de la estructura del ojo donde se manifiesta esta enfermedad.

La divulgación de información sobre nuevos tratamientos y sobre investigación científica reciente se refleja en varias de las publicaciones, lo que deja entrever un sentimiento de esperanza y expectativas positivas hacia la detención o reversión del proceso degenerativo que lleva a la pérdida de la visión. Estas expresiones emotivas se hacen presentes, sobre todo, en los comentarios de personas con ES y sus padres: "vamos a contrarreloj… lloré de la emoción… si funciona [en pruebas en el extranjero], por qué no podemos???!!!" (Post n.º 45 de Facebook) y "estoy deseando que llegue el día en que pueda decir que mi hijo está participando en un estudio de esos" (Post n.º 17 de Facebook). También se difunden desde el GVS cursos sobre investigaciones y tratamientos nuevos, demostrando de nuevo el interés y la esperanza hacia todo lo que rodea a este tema.

**(T6) tenemos una discapacidad visual, pero somos más que eso**

Para las personas afectadas por la ES, la discapacidad visual es solo uno más de los aspectos de su vida, pero no lo que las define. Los discursos recogidos en GVS tratan la discapacidad, más que como una condición o propiedad del paciente, como una característica que se manifiesta en la relación de este con su contexto. Tomando la posición del Estatuto de Defesa dos Direitos da Pessoa com Deficiência publicado en 2015, la discapacidad "no está en las personas sino en las barreras debidas a las actitudes y al ambiente que impiden el ejercicio efectivo pleno de los derechos de todos, en igualdad de condiciones y oportunidades" (Enlace del post n.º 32 de Facebook). Esta afirmación parte de la premisa de que la igualdad de oportunidades constituye un derecho fundamental, para el que la baja visión o la ceguera no deben suponer barreras.

En palabras de Sony, la persona con discapacidad vive hoy una situación más favorable para el ejercicio pleno de su libertad. Según él, esto se debe a los avances en lo que respecta a las tecnologías de apoyo, a la movilidad y a la propia actitud y percepción de la sociedad. Siguen existiendo diversas barreras, pero "hoy las personas con discapacidad pueden hacer realidad sus sueños, viajan, planifican su vida, gastan, compran. ¡Eso está muy bien!" (Vídeo del post n.º 13 de Facebook).

Se encontraron muchas declaraciones sobre cómo estas personas construyen su identidad a partir de distintas actividades, lo que les hace superar la concepción simple de portadores de una discapacidad visual. Ejemplos como el de Luciano con los deportes, Lúcia con el maquillaje, Marina con el ballet o Milka con la masoterapia, demuestran que cada uno de ellos es capaz de desarrollar habilidades, incluso con poca capacidad de visión. La masoterapia aparece como una práctica en la que la persona con ES puede tener un enorme potencial, ya que se ejerce mediante sensibilidad táctil en gran medida (Vídeo de YouTube de 28 de enero de 2017). Los deportes paraolímpicos se presentan también en GVS como un campo de actuación de la persona con discapacidad, lo que se plasma en declaraciones y entrevistas de nadadores, levantadores de peso, y judocas (Vídeos de los posts n.º 52 y 68 de Facebook).

**(T7) convivimos con ciertos problemas y equívocos por la forma en que vemos las cosas**

Vivir con baja visión es vivir de modo constante con equívocos y problemas. Al ver lo suficiente como para caminar por la calle, pero no ser capaz de leer un texto impreso en un papel o reconocer los rostros de la gente puede llevar a situaciones complicadas. Algunas de esas situaciones incómodas se deben a la dificultad de los demás para entender que la persona con enfermedad de Stargardt a veces necesita ayuda o condiciones especiales, a pesar de ser capaz de mantener una existencia anónima como videntes de rango normal. Las declaraciones de Barbara hacen hincapié en esos problemas por no tener un status socialmente reconocido y legitimado:

> Tener baja visión es ser ciego en un mundo de videntes y vidente en un mundo de ciegos. Es que te miren mal cuando estás usando el móvil y estás sentado en un banco preferente. Es que no se acuerden de ti cuando se habla de accesibilidad de persona con discapacidad visual. Es no recibir ayuda porque ves. Es tener que usar el bastón para que todo el mundo lo sepa y, aún así, que se te considere un impostor. Es que se olviden de ti los compañeros y los profesores porque actúas como los demás. Pero tiene un lado bueno: ves. (Vídeo "O que é baixa visão", indicado en el post n.º 49 de Facebook)

Paulo también apunta este sentimiento de no pertenencia a ningún grupo cuando dice, en tono indignado: "[…] sabe esa sensación de saber que alguien le está observando? […] la gente quiere estar segura de si soy o no invidente". Preguntándose el porqué de que las personas asuman esa actitud de desconfianza, el propio Paulo subraya para las situaciones en las que no ha sido ayudado ante un imprevisto o accidente: [¿esa persona que está ahí de pie] qué está esperando? ¿a que yo tropiece y caiga para demostrarle que es verdad que no veo bien?¨ (Vídeo del post n.º 13 de Facebook). Otro de los testimonios recogidos señala la incomodidad que produce la desconfianza de los demás, lo que afecta a la autoestima y a la concepción de la valía personal. Según Milka, esa desconfianza hizo que "en algunos momentos, llegué a dudar de si era realmente capaz de hacer determinadas cosas" (Vídeo de YouTube del 21 de enero de 2017).

Incluso en la interacción con personas conocidas, que saben de la discapacidad visual del enfermo de ES, también se dan las situaciones de incomprensión momentánea sobre la capacidad de ver bien. La intención de la persona con ES de omitir su dificultad en la visión, expresada por Marina, Gabriela y Luciano, demuestra lo cansado que puede llegar a ser estar siempre explicando qué cosas puede ver y cuáles no. Por tanto, la alternativa es fingir que ve bien. "[Gabriela:] no conseguimos distinguir los rasgos faciales. Pero …[Marina:] miramos […] y fingimos que vemos todo. [Gabriela:] […] ¡Hacemos eso, es verdad!" (Vídeo de YouTube del 2 de marzo de 2016). Lo mismo sucede cuando alguien les enseña algo en un móvil o en un cartel, tal y como relata Luciano:

> *Otra cosa que solemos hacer es engañar a la gente […] Como nos miran y no ven nada particular en nuestros ojos […] para ellos somos personas normales. Entonces, esa persona llega toda contenta y nos dice "mira esto" y nos muestra una foto, nos manda leer un mensaje. Ahí finges que lees, sueltas una carcajada y dices "¡madre mía!", haces el paripé, merecedor de Oscar. Esa persona se va toda contenta, pero tú no viste nada, no entendió ni papa.* (Vídeo de YouTube del 9 de julio de 2017).

Estas situaciones complicadas y surrealistas, pasado algún tiempo, se incorporan a un conjunto de historias graciosas vividas. Denominadas *gafies* o *patetadas* por Luciano, él mismo las presenta como habituales para la persona con ES. En un vídeo dedicado a estas situaciones ridículas, hace un recorrido por este tipo de vivencias, como son: confundir maniquíes en las puertas de las tiendas y conversar con ellos o equivocarse de coche y entrar en el de un extraño creyendo que es su mujer (Vídeo de YouTube del 9 de julio de 2016).

Este sentido del humor, que permite convertir una dificultad o una limitación en algo divertido, no siempre está presente en los testimonios del grupo. Pese a ello, cuando una persona con esta enfermedad se encuentra con otra que también lo padece, las bromas relacionadas con la baja visión son habituales en sus conversaciones. Un ejemplo de ello es la presentación de sí mismo que hace Roberto al inicio de un curso. En tono irónico, se aprovecha de las dificultades para verlo que tiene el público y dice: "[…] mido 1,99m., peso 40 kilos, tengo el pelo rubio y largo hasta los hombros, y creo que guapo, porque hasta este momento [ustedes] no han mirado a nadie más" (Vídeo de YouTube de 27 de marzo de 2017).

**(T8) nos reconocemos como comunidad**

Las cuestiones genéticas (biológicas) que determinan la enfermedad de Stargardt, las consecuencias que derivan a la baja visión (aspectos funcionales) y la similitud en cuanto a experiencias (aspectos vivenciales) de los miembros del GVS crean una noción de familiaridad, de cercanía y de comunidad. Según revelan los administradores del grupo y por algunos otros participantes, ha terminado acuñándose la expresión "familia Stargardt" para calificar lo que se establece entre sus pares. Ello denota el sentimiento de pertenencia a un grupo en el que la enfermedad es el lazo biológico en el que se reconocen mutuamente. El GVS promueve diversos actos y eventos de confraternización e intercambio de información, de manera que se facilite el contacto cara a cara de sus miembros (Post n.º 18 de Facebook).

Según la actividad formativa propuesta por uno de los administradores del grupo, todo ello va más allá de una trayectoria en que las personas con ES y sus familiares buscan informaciones sobre esta patología e intercambian experiencias. Con la extensión de internet en Brasil, se han creado blogs y comunidades virtuales desde el año 2000, destacando el grupo de discusión Stargardt de Yahoo, que llegó a reunir a cerca de 200 participantes de todo el país (Vídeo de YouTube del 27 de marzo de 2017). Una vez que integran estas comunidades, aumentan las iniciativas de intereses comunes a sus miembros, como la divulgación científica y el acceso a investigaciones y tratamientos, la difusión y defensa de los derechos legales de las personas con discapacidad y la consolidación de canales de comunicación entre pacientes y sus familiares.

El uso de *hashtags* como *#JuntosSomosMaisFORTES* y *#deficientesVisuais* en algunas de sus publicaciones denota la intención del GVS de articularse como dinamizador social comprometido con los intereses comunes de las personas con discapacidad visual. Del mismo modo, también se involucra en los intereses compartidos por otras asociaciones y comunidades. Entre ellas se mencionan: Retina Brasil y sus delegaciones regionales, y el Mobimento Brasileiro de Mulheres Cegas e com Baixa Visão (MBMC). Entre las iniciativas sociales defendidas por el GVS cabe destacar el apoyo al Proyecto de Ley de la Cámara n.º 56/2016, de Política Nacional sobre Enfermedades Raras, que propone, entre otros aspectos, que se proporcione gratuitamente medicación a los enfermos de este tipo de enfermedades (Post n.º 39 de Facebook).

## 4 Conclusiones

Las comunidades virtuales y los grupos en redes sociales son formas en que se plasman las nuevas modalidades de comunicación interactiva en salud. El aumento del número de usuarios, el tiempo conectado y la predisposición de los participantes por discutir sobre temas y asuntos de salud, hábitos de riesgo, miedos e inexactitud de las informaciones hace que el acceso a la información aumente. Además, "la solicitud de apoyo también hace que crezca la dimensión de las informaciones, lo que también aumenta la audiencia potencial" (Silveira, Costa & Lima, 2012). Es precisamente en este nuevo entorno en el que estas vías de comunicación pueden contribuir a aumentar el espectro de las relaciones personales, siempre, eso sí, que los usuarios no las conciban solo como una herramienta para perpetuar las relaciones sociales ya existentes sino también como una forma de establecer conexiones nuevas. Yendo un paso más allá, como plantean Martiniello et al (2012), es fundamental llamar la atención sobre la necesidad de estar presentes y visibles en las redes, puesto que mientras mayor sea su protagonismo, y más consideración tengan como grupo, más se extenderán las aplicaciones y características que las harán usables y adecuadas para determinados colectivos, como son las personas con baja visión.

Cabe destacar, como uno de los hallazgos del estudio, que esta comunidad se muestra transparente y proclive a difundir las características de su modo de vida en las redes de forma directa y natural, algo que en cierto modo supone un cambio de percepción de estos instrumentos con respecto a las conclusiones que se han puesto de manifiesto en estudios anteriores, en los que las personas con problemas de visión se mostraban más reacias a hacer preguntas sobre estos temas, tanto por la falta de respuesta como por la posibilidad de aparecer como personas dependientes frente al resto de sus contactos (Brady et al, 2013). Frente a esa apreciación, y en la línea de lo planteado por Wu y Adamic (2014) aplicado a personas con dificultades visuales, los resultados indican que las personas con ES sí valoran estos medios y redes sociales como un cauce para hacerse oír, plantear sus preocupaciones y obtener ayuda y atención, lo que termina derivando en mayor visibilidad. De hecho, ese mismo estudio comprobaba que este tipo de contenidos recibían de media mayor tasa de respuesta (comentarios y "me gusta" en Facebook) que los que no eran tan personales.

Los resultados obtenidos demuestran que la comunicación interactiva en salud, la relación entre ejes conceptuales cognitivos, sociales y tecnológicos están presentes en la construcción identitaria de esta comunidad (Caran & Biolchini, 2016). Padecer la enfermedad de Stargrardt va mucho más allá de la cuestión patológica en sí misma, sino que tiene que ver con la vivencia de experiencias complicadas parecidas en el día a día, con la indignación por la desigualdad de oportunidades, con las esperanzas en nuevos tratamientos y en las tecnologías de apoyo. Estas realidades vividas se comparten en el GVS y van construyendo el discurso intersubjetivo de "qué significa ser una persona con ES" y de "qué no es una persona con ES". Esta construcción de la identidad toma forma a través de un proceso de comunicación interactiva basada en la negociación de elementos simbólicos, alternando los momentos de acuerdo, desacuerdo y consenso (Gajaria, Yeung, Goodale & Charach, 2011). En definitiva, y tal y como adelantaban Morel y Villalobos (2011), estas nuevas formas de comunicación ayudan a generar un 'constructo social de la discapacidad' positivo y diferente, capaz de valorar la diversidad y respetar al sujeto.

## 4 Referencias